\documentclass[pra,aps, english, twocolumn, hyperref]{revtex4}

\usepackage{graphicx}

\usepackage{amssymb, amsmath, color, rotating}
\begin{document}

\title{Spin dynamics in a spin-orbit coupled Fermi gas}

\author{Stefan S. Natu}

\email{snatu@umd.edu}

\affiliation{Condensed Matter Theory Center and Joint Quantum Institute, Department of Physics, University of Maryland, College Park, Maryland 20742-4111 USA}

\author{S. Das Sarma}

\affiliation{Condensed Matter Theory Center and Joint Quantum Institute, Department of Physics, University of Maryland, College Park, Maryland 20742-4111 USA}

\begin{abstract}
We study the dynamics of a non-degenerate, harmonically trapped Fermi gas following a sudden ramp of the spin-orbit coupling strength using a Boltzmann equation approach. In the absence of interactions and a Zeeman field, we solve the spin-orbit coupled Boltzmann equation \textit{analytically}, and derive expressions for the phase-space and temporal dynamics of an arbitrary initial spin state. For a fully spin polarized initial state, the total magnetization exhibits collapse and revival dynamics in time with a period set by the trapping potential. In real space, this corresponds to oscillations between a fully polarized state and a spin helix. To make predictions relevant to  current experiments on spin-orbit coupled Fermi gases, we then numerically study the dynamics in the presence of an additional  momentum independent Zeeman field. We find that the spin helix is robust for weak magnetic fields but disappears for stronger field strengths. Finally, we explore the spin dynamics in the presence of interactions and find that weak interactions \textit{enhance} the amplitude of the spin helix.  
\end{abstract}
\maketitle

\section{Introduction}  The physics of spin-orbit coupling is at the heart of fundamental phenomena such as the spin Hall effect \cite{hirsch, zutic} as well as practical devices such as the spin transistor \cite{dattadas}. Key to these developments in the field of spintronics is the understanding of how parameters such as the spin-orbit coupling, interaction, disorder, and geometry separately, as well as collectively influence spin dynamics \cite{bernevig, halperin}. The creation of low temperature atomic and molecular spin-orbit coupled Bose and Fermi gases has paved the way for studying this physics in a setting where these parameters are well characterized \cite{nistexpt, spielman2, zwierleinsoc, chinasoc, spielmanferm}. Furthermore, a tool largely unique to ultra-cold gases is the ability to induce the spin-orbit coupling \textit{dynamically}, thereby enabling the study of \textit{out of equilibrium} physics in these systems. In this paper, we solve the dynamics of a non-interacting non-degenerate Fermi gas following a sudden ramp of the spin-orbit coupling strength. The resulting out-of-equilibrium dynamics is rich \cite{stanescu, anderson, sherman}: coherence inherent in cold atomic gases, but almost always absent in solid state systems leads to collapse and revival of the total magnetization \cite{stanescu}. In real space, this is manifested by the spontaneous appearance of a helical spin texture which is the analogue of the persistent spin helix observed in two-dimensional electron gases \cite{bernevigsu2, spinhelix}. 

A further advantage of cold atomic systems is the ability to use magnetic fields to control the interactions between the atomic states \cite{blochreview, ketterlereview}. This, particularly in the presence of spin-orbit coupling and s-wave superfluidity, may enable experimentalists to realize novel states of matter with topological properties whose excitations exhibit non-Abelian statistics \cite{sau, zhang}. In addition to their fundamental importance, such topological states of matter may serve as a platform for fault tolerant quantum computation \cite{kitaev}. However a key challenge in observing this physics in ultra-cold gases is the inherent difficulty in attaining sufficiently low temperatures needed to realize these topological states. In contrast, the non-equilibrium dynamics we study here occurs at high temperatures and can be observed in \textit{current} experiments. We numerically study the spin dynamics of a fully polarized, \textit{weakly} interacting Fermi gas using a collisionless Boltzmann equation. We find large amplitude spin waves analogous to those previously observed in dilute spin polarized Hydrogen \cite{bashkinspin, laloe, mullin, castaing, johnsonspin} and more recently  in ultra-cold Bose and Fermi gases \cite{thomasspinwave, cornellspinwave, natuspindyn, eckardt, laloe2, oktel}. Our numerical study of spin dynamics in the collisionless spin-orbit coupled Fermi gas complements a recent, linear response study on a homogeneous system by Tokatly and Sherman \cite{sherman}. We point out that the ultra-low temperature degenerate version of our system (with spin-orbit coupling and large Zeeman splitting) would manifest topological superfluidity in the presence of ordinary s-wave superfluidity induced by suitable Feshbach resonance \cite{zhang}.

We show that spin-orbit coupling, when combined with the long coherence times inherent to cold atomic and molecular gases, leads to surprising dynamical phenomena even at high temperatures where the cold gas is not necessarily quantum degenerate. Furthermore the non-degenerate gas is an ideal conceptual starting point for studying how interactions influence the spin dynamics in the presence of spin-orbit coupling.

The effects predicted in our work are unlikely to be of much experimental significance in solid state spin-orbit coupled systems because of strong disorder and decoherence intrinsically present in these systems as well as the ultra-fast time scales for spin relaxation \cite{dperel}.  But in cold atomic and molecular systems, our proposed physics could be studied in \textit{existing} laboratory systems.  In principle, the physics we predict should be present in both atomic/molecular Fermi/Bose gases since it is an intrinsically high temperature phenomenon, but in light of recent experiments \cite{zwierleinsoc, chinasoc, spielmanferm}, we will discuss our theoretical details using the atomic Fermi gas as the representative system of study.

This paper is organized as follows: In Section II, we describe our system and derive the collisionless Boltzmann equation for a two-component Fermi gas in a $2$D harmonic trap in the presence of spin-orbit coupling. In the subsequent sections, we choose an initial state which is a stationary state of the Hamiltonian in the absence of spin-orbit coupling. We then drive the system out-of-equilibrum by suddenly turning on the spin-orbit coupling and study the resulting dynamics. In Section III, we solve the Boltzmann equation in the absence of interactions, and in Section IV, we consider the effect of interactions on the spin dynamics. We summarize our results in Section V.

\section{Setup}  In the absence of spin-orbit coupling, the Hamiltonian for two hyperfine states of a Fermi gas can be expressed as a sum of single particle and two-body interaction terms:
\begin{equation}
{\cal{H}} = {\cal{H}}_{\text{s}} + {\cal{H}}_{\text{int}}
\end{equation}
where the single particle Hamiltonian is composed of a kinetic term and a potential term arising from the external trapping potential
\begin{equation}\label{kinham}
{\cal{H}}_{\text{s}} = \sum_{i = \uparrow,\downarrow}~\int d^{3}\textbf{r}~\psi^{\dagger}_{i}\Bigl(-\frac{\hbar^{2}\nabla_{\textbf{r}}^{2}}{2m} + U(\textbf{r})\Bigr)\psi_{i} \\\nonumber
\end{equation}
where $\psi_{i}$ denotes the fermionic annihilation operator for a particle in hyperfine state $\{\uparrow, \downarrow\}$ and mass $m$. Here $U(\textbf{r})$ refers to the external trapping potential, which we assume to be cylindrically symmetric  $U(\textbf{r}) = \frac{1}{2}m\omega_{r}^{2}(x^2+y^2) + \frac{1}{2}m\omega_{z}^{2}z^2$, where $\omega_{r}$ and $\omega_{z}$ are the trapping frequencies in the radial and longitudinal directions respectively. As the relevant spin-orbit physics is two-dimensional, we assume a quasi-two dimensional, pancake geometry, obtained by tight confinement in the longitudinal ($z$) direction ($\omega_{z} \gg \omega_{r}$). For simplicity, we assume that  both the atomic states experience identical trapping potentials, but all our calculations can be readily extended to include more general trapping potentials realized in experiments.

At the ultra-cold temperatures realized in these experiments, the dominant contribution to scattering comes from the s-wave channel. The interaction Hamiltonian therefore takes the simple form of a contact interaction:
\begin{equation}\label{intham}
{\cal{H}}_{\text{int}} = \frac{g}{2}\int d\textbf{r}~\psi^{\dagger}_{\uparrow}(\textbf{r})\psi^{\dagger}_{\downarrow}(\textbf{r})\psi_{\downarrow}(\textbf{r})\psi_{\uparrow}(\textbf{r})
\end{equation}
with interaction strength $g = 4\pi \hbar^{2}a/m$ where $a$ denotes the s-wave scattering length. A key advantage of the fermonic experiments \cite{zwierleinsoc, chinasoc, spielmanferm} is the ability to use magnetic fields to tune the interactions between different internal states, via  a Feshbach resonance \cite{ketterlereview, blochreview}. Here we work in the weakly interacting regime, which can be realized by working near the zero crossing of the Feshbach resonance.  

The spin-orbit Hamiltonian containing terms linear in momentum takes the form:
\begin{equation}\label{hsoc} 
{\cal{H}}_{\text{SOC}} = \alpha(\sigma_{x}p_{y} - \sigma_{y}p_{x}) + \beta(\sigma_{x}p_{x} - \sigma_{y}p_{y})
\end{equation}
where the first term is the Rashba contribution and the second term is the Dresselhaus contribution, parametrized by the coupling constants $\alpha$ and $\beta$ respectively. Here $\sigma_{x}, \sigma_{y}$ and $\sigma_{z}$ denote Pauli matrices. 

In the ultra-cold gas setting, spin-orbit coupling is generated by using a pair of Raman beams to drive transitions between two hyperfine states of an atom, while simultaneously imparting a momentum kick \cite{nistexpt, spielman2, zwierleinsoc, chinasoc}. The resulting spin-orbit Hamiltonian takes the Rashba equal Dresshelhaus form $\alpha = \beta$, where the magnitude of $\alpha$ is determined by the wave-length of the Raman beams and the angle at which they intersect. The Raman beams also produce a momentum independent Zeeman field,  ${\cal{H}}_{Z} = -\hbar\Omega_{R}\sigma_{z}$, where $\Omega_{R}$ is the strength of the Raman coupling, and is proportional to the intensity of the Raman lasers. As we show below, this term plays an important role in the dynamics. The scheme described here was first successfully demonstrated in experiments at NIST using bosonic $^{87}$Rb \cite{nistexpt, spielman2}. Recently, a similar scheme was used to generate spin-orbit coupling in a Fermi gas of $^{6}$Li and $^{40}$K \cite{chinasoc, zwierleinsoc, spielmanferm}. 

As we are primarily motivated by current experiments, we limit ourselves to the case $\alpha = \beta $ in Eq.~\ref{hsoc}. In this limit, the spin-orbit Hamiltonian can be diagonalized by independently rotating the momentum co-ordinate and performing a global spin rotation. We will denote the diagonal basis as $\{ \psi_{+}, \psi_{-}\}$. Throughout, we will use both the diagonal basis ($\{+, -\}$) and the pseudo-spin basis ($\{\uparrow, \downarrow\}$) corresponding to the original hyperfine states, depending on the context. The generalization to $\alpha \neq \beta$ is straightforward within our formalism, and will be the subject of a future work \cite{jurajinprep}. 

All of the work described here is in the \textit{non-degenerate} limit. The only requirement on the temperature $T$ is that it should be smaller than the detuning energy between the magnetic sublevels, so that the gas can be described by a two-level (pseudo-spin $\frac{1}{2}$) system. This is readily accomplished as the splitting between hyperfine levels is much larger than the Fermi energy for typical densities \cite{chinasoc}.  

The physics of the weakly interacting gas is dominated by coherent mean-field dynamics which occurs on a timescale $\tau_{\text{mf}} \sim (a n_{0}/m)^{-1}$ which is much faster than the timescale for energy exchanging collisions $\tau_{\text{coll}} \sim (4\pi a ^2n_{0}v)^{-1}$ ($v$ is the characteristic velocity of the particles and $n_{0}$ is the density). For a trapped gas of $N \sim 10^{5}$ $^6$Li atoms at a temperature $T \sim 10^{-6}$K, the collisionless limit ($\tau_{\text{mf}} \ll \tau_{\text{coll}}$) corresponds to scattering lengths $a \sim 10 a_{\text{B}}$, where $a_{\text{B}}$ is the Bohr radius. The recent experiment of Cheuk \textit{et al.} is already in this regime \cite{zwierleinsoc}, while the experiment of Wang \textit{et al.} \cite{chinasoc} has stronger interactions of $a \sim 200a_{\text{B}}$, which can be tuned near zero using a Feshbach resonance \cite{ketterlereview}. 

Mathematically, the weakly interacting gas can be described using a collisionless Boltzmann equation. Following Ref. \cite{kadbaym}, we use the Heisenberg equations for $\psi_{\sigma}$(r, t) to derive the equations of motion for the
spin dependent Wigner function
\begin{eqnarray}\label{wigfuns}
\overleftrightarrow{\textbf{F}} &=& 
\left(\begin{array}{cc} 
f_{\uparrow\uparrow}(\textbf{p}, \textbf{R}, t)&f_{\uparrow\downarrow}(\textbf{p}, \textbf{R}, t) \\ 
f_{\downarrow\uparrow}(\textbf{p}, \textbf{R}, t)&f_{\downarrow\downarrow}(\textbf{p}, \textbf{R}, t) \end{array}\right)\\
\nonumber
f_{\sigma \sigma^{'}}(\textbf{p}, \textbf{R}, t) &=& \int d \textbf{r} e^{i \textbf{p}\cdotp \textbf{r}} \langle \psi^{\dagger}_{\sigma}(\textbf{R} - \frac{\textbf{r}}{2}, t) \psi_{\sigma^{'}}(\textbf{R} + \frac{\textbf{r}}{2}, t) \rangle, 
\end{eqnarray}
which is the quantum analogue of the classical distribution function. Here $\textbf{p}$ represents the momentum, $\textbf{r} = \textbf{r}_{1} - \textbf{r}_{2}$ is the relative coordinate and $\textbf{R} = \frac{\textbf{r}_{1}+\textbf{r}_{2}}{2}$ is the center of mass coordinate.  

The diagonal components of $\overleftrightarrow F$ can be integrated in momentum to give the respective spin densities $n_{\sigma\sigma}(\textbf{R}, t) =  \langle \psi^{\dagger}_{\sigma}(\textbf{R}, t) \psi_{\sigma}(\textbf{R},  t) \rangle =  \int \frac{d \textbf{p}}{(2\pi)^{3}}f_{\sigma\sigma}(\textbf{p}, \textbf{R}, t)$, while the off-diagonal components correspond to quantum coherences that are absent in a classical model of a spin-$\frac{1}{2}$ gas. 

For the pancake geometry considered here, all the relevant dynamics is two dimensional. Assuming thermal equilibrium in the longitudinal direction, we  decompose the Wigner function in the radial and axial directions as $f_{\sigma\sigma^{'}}(\textbf{p}, \textbf{R}, t) = f_{\sigma\sigma^{'}}(p_{r}, r, t)f(p_{z}, z)$ where $f(p_{z}, z) = e^{-\beta(p_{z}^{2}/2m + \frac{1}{2}m\omega^{2}_{z}z^2)}$ where $\beta = 1/k_{B}T$.

We obtain a two dimensional density $n^{\text{2D}}_{\sigma\sigma^{'}}(r, t) = \int dp_{r} f_{\sigma\sigma^{'}}(p_{r}, r, t)\int dp_{z}f(p_{z}, z)$ by integrating out the longitudinal co-ordinate. Averaging over the $z-$direction, we obtain an effective quasi $2$D Boltzmann equation:
\begin{eqnarray}\label{boltzmann}
\partial_t\overleftrightarrow{\textbf{F}} + \frac{\textbf{p}}{m}\cdotp\nabla_{\textbf{r}}\overleftrightarrow{\textbf{F}} - \nabla_{\textbf{r}}U\nabla_{\textbf{p}}\overleftrightarrow{\textbf{F}} = i [\overleftrightarrow{\textbf{V}}, \overleftrightarrow{\textbf{F}}] + \hspace{15mm} \\\nonumber \frac{1}{2}\{\nabla_{\textbf{r}}\overleftrightarrow{\textbf{V}},\nabla_{\textbf{p}}\overleftrightarrow{\textbf{F}}\} + 
i \alpha p_{+} [\sigma_{+}, \overleftrightarrow{\textbf{F}}] + \frac{\alpha}{2} \{\sigma_{+}, \nabla_{r_{+}}\overleftrightarrow{\textbf{F}}\}
\end{eqnarray}

where $p_{+} = p_{x}+p_{y}$, $\sigma_{+} = \sigma_{x}+\sigma_{y}$, and the interaction potential $\overleftrightarrow{\textbf{V}}$ is \cite{natuspindyn}:
\begin{equation}\label{intpot}
\overleftrightarrow{\textbf{V}} =  \left(\begin{array}{cc} g^{\text{2D}}n_{\downarrow\downarrow} - \hbar\Omega_{R}&-gn_{\uparrow\downarrow}\\ -gn_{\downarrow\uparrow}&gn_{\uparrow\uparrow} + \hbar\Omega_{R}
\end{array}\right)
\end{equation} 
where $g^{\text{2D}}$ is an effective two-dimensional interaction strength given by $g^{\text{2D}} = 2\sqrt{\pi}\hbar^{2}a/(m\Lambda_{\text{th}})$, where $\Lambda_{\text{th}} = \sqrt{2\pi\hbar^{2}/mk_{B}T}$. The diagonal components in the interaction matrix arise from forward scattering (Hartree) while the off-diagonal terms arise from exchange interactions (Fock). Commutators and anti-commutators are denoted by $[,]$ and $\{,\}$ respectively. 

The single particle limit of Eq.~\ref{boltzmann} was derived by Mishchenko and Halperin \cite{halperin}, who used this approach to study the transport properties of a $2$D electron gas. Here we generalize the Boltzmann equation to the include effect of $\overleftrightarrow V$ on the phase space and spin space evolution of the Wigner function.

 In general, Eq.~\ref{boltzmann} is a non-linear matrix equation which has to be solved numerically. Below we assume an initial state which is a stationary state of the Hamiltonian in the absence of spin-orbit coupling ($\alpha = \Omega_{R} = 0$). We then suddenly turn on the Raman coupling, and investigate the resulting out-of-equilibrium dynamics. 
 
 \section{Non-interacting limit}  
\subsection{$\Omega_{R}  = 0$}
The non-interacting limit in ultra-cold Fermi gases is achieved by working at the zero crossing of a Feshbach resonance. As the magnetic fields corresponding to the zero crossing of the interactions are typically small \cite{zwierleinsoc}, we do not expect the Raman couplings to deviate appreciably from their zero field values \cite{weimueller}. We first consider the case where upon switching on the Raman coupling, the spin-orbit coupling ($\alpha$) is non-zero, but the Zeeman term $\Omega_{R} = 0$.  While this scenario does not correctly model the present experiments, in this limit the Boltzmann equation is exactly soluble for an arbitrary initial spin state, thus serving as a conceptual starting point. We remark that although we only consider the non-degenerate limit here, our results can be readily generalized to temperatures below the Fermi temperature. 

In order to proceed, we introduce dimensionless position and momentum coordinates $\tilde r = r/r_{\text{trap}}$ and $\tilde p = p~\sqrt{2\pi}\hbar/\Lambda_{\text{th}}$, where $r_{\text{trap}} = \sqrt{\hbar/m\omega_{r}}$ is the characteristic length scale of motion in the trap and and $\Lambda_{\text{th}} = \sqrt{2\pi\hbar^{2}/mk_{B}T}$ is the thermal deBroglie wavelength. We normalize time in units of the radial trapping frequency $\tilde t = t/\omega_{r}$. We also introduce a parameter $\eta = \sqrt{\hbar\omega_{r}/k_{B}T}$. As the spin-orbit Hamiltonian only couples to momentum in the $p_{x}+p_{y}$ direction, it suffices to consider the evolution of the distribution in the $p_{+} = p_{x}+p_{y}$ and $r_{+} =x+y$ directions of phase space. 

Consider an arbitrary initial spin state given by the Wigner distribution function: $\overleftrightarrow{\textbf{F}}(\tilde p_{+}, \tilde r_{+}, t=0) = e^{-\frac{1}{4}(\tilde p^{2}_{+}+ \eta^{2}r^2_{+})}\overleftrightarrow f$ where $\overleftrightarrow f$ is a $2\times2$ matrix corresponding to the initial spin state. We omit the $p_{-}, r_{-}$ directions for now as they have no dynamics. In the absence of spin-orbit coupling or interactions ($\alpha = \Omega_{R} = g^{\text{2D}} = 0$), the initial state is stationary.

Next, we express the Boltzmann equation in the diagonal basis by performing the global transformation $\overleftrightarrow F \rightarrow B^{\dagger} \overleftrightarrow f B$ where the unitary matrix $B = \left(\begin{array}{cc}\frac{1}{\sqrt{2}}&\frac{-1+i}{2} \\ 
\frac{1+i}{2}&\frac{1}{\sqrt{2}} \end{array}\right)\\$ rotates $\sigma_{x} + \sigma_{y}$ to $\sqrt{2}\sigma_{z}$. In the rotated basis, the spin $+$ and $-$ components evolve independently, and the problem reduces to a single particle problem in the presence of a \textit{momentum dependent} magnetic field. 

\begin{figure}
\begin{picture}(100, 165)
\put(-60, -10){\includegraphics[scale=0.5]{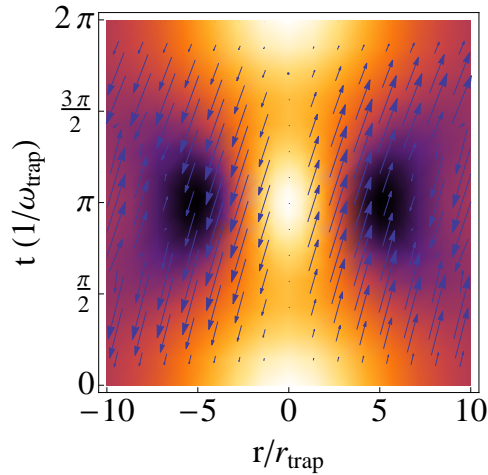}}
\end{picture}

\caption{\label{nonintfig} Time evolution of the longitudinal (density plot) and transverse magnetization (arrows) in the $r_{+} = x+y$ direction, following a sudden ramp of the spin-orbit coupling strength. The Zeeman term is set to zero here ($\Omega_{R} =0$). Brighter colors indicate positive magnetization ($\uparrow$) and darker colors indicate negative magnetization ($\downarrow$). The arrows indicate the direction of the magnetization in the $x-y$ plane, and the length of the arrows indicate the magnitude of the transverse spin normalized to the total spin. At $t=0$ all the spins are pointing in the $z$ direction (all atoms in the $\uparrow$ state). spin-orbit coupling causes the atoms to precess in time, and at half the trapping period a spin helix is produced. The wave-vector of the helix at $t = \pi/\omega_{r}$ depends only on the spin-orbit interaction and is $\lambda_{\text{sh}} = \pi~r_{\text{trap}}/2\tilde\alpha$, where $r_{\text{trap}} = \sqrt{\hbar/m\omega_{r}}$. To clearly illustrate this effect, we choose a weak spin-orbit coupling of $\tilde\alpha = \sqrt{2}\alpha \sqrt{m/\hbar\omega_{r}} = 0.125$ and $\eta =  \sqrt{\hbar\omega/k_{B}T} = 0.25$ in these simulations.  The initial state is recovered after $t = 2\pi/\omega_{r}$.}
\end{figure}

The dynamics in phase space can now be solved by making the following ansatz for the diagonal components of the rotated spin density matrix:
\begin{eqnarray}\label{diagansatz}
F_{++/--}(\tilde p_{+}, \tilde r_{+}, \tilde t) = A_{+/-}e^{-\frac{1}{4}\{(\tilde p_{+} \mp a(\tilde t))^2 + \eta^2(\tilde r_{+} \mp b(\tilde t))^2\}}
\end{eqnarray}
where $a(0) = b(0) = 0$. The coefficients $A_{+/-}$ are the diagonal matrix elements of the spin density matrix $\overleftrightarrow f$ after rotation into the $\{+, -\}$ basis. 

Substituting the ansatz of Eq.~\ref{diagansatz} into Eq.~\ref{boltzmann} we find $a(\tilde t) = \tilde\alpha \eta (\cos(\tilde t)-1)$ and $b(\tilde t) = \tilde\alpha \sin(\tilde t)$, where we have introduced a dimensionless, spin-orbit coupling constant $\tilde\alpha = \sqrt{2}\alpha \sqrt{m/\hbar\omega_{r}}$, which parametrizes the strength of the spin-orbit coupling relative to the trapping potential. Thus the rotated spin densities simply perform oscillations in real and momentum space with an amplitude set by the strength of the spin-orbit interaction and period set by the trap frequency. 

Similarly, one can solve for the dynamics of the off-diagonal components to find:
\begin{eqnarray}\label{offdiageq}
F_{+-}(\tilde p,_{+} \tilde r_{+}, \tilde t) = A_{+-}e^{\frac{-4\tilde\alpha^{2}(1-\cos(\tilde t))}{\eta^{2}}} \times \\\nonumber e^{-\frac{1}{2}\{(\tilde p_{+} -2 i \tilde\alpha/\eta\sin(\tilde t))^2 + \eta^2(\tilde r_{+} - 2i\tilde\alpha/\eta^2(1-\cos(\tilde t)))^2\}}\\\nonumber
\end{eqnarray}
where $A_{+-}$ is the off-diagonal matrix element of the spin density matrix after rotation into the $\pm$ basis. The dynamics of $F_{-+}$ is obtained by replacing $\tilde\alpha \rightarrow -\tilde\alpha$ in Eq.~\ref{offdiageq}. Unlike the diagonal components, the magnitudes of the off-diagonal components are \textit{not} conserved and oscillate in time. 

The corresponding spin densities are found by integrating the above expressions for the Wigner functions in momentum space. By rotating the diagonal basis back into the hyperfine basis \{$\uparrow, \downarrow$\}, one obtains the dynamics of an arbitrary initial spin state. 

To illustrate the role of quantum coherence, we consider the dynamics of a fully polarized initial state corresponding to all particles in the $\uparrow$ state. We study the dynamics of the longitudinal magnetization density and the total magnetization:
\begin{eqnarray}\label{magmagdens}
m_{z}(\textbf{r}, t) = \int d\textbf{p}~\Big(f_{\uparrow\uparrow}(\textbf{p}, \textbf{r}, t) -f_{\downarrow\downarrow}(\textbf{p}, \textbf{r}, t)\Big) \\\nonumber
M(t) = \int d\textbf{r}~m_{z}(r, t) \hspace{30mm}
\end{eqnarray}

The zero temperature dynamics of the total magnetization for this initial state was considered previously by Stanescu, Zhang and Galitski \cite{stanescu}. By exactly solving for the quantum dynamics in a trap, they demonstrated that the total magnetization exhibits collapse and revival dynamics, and produced analytic formulas for the total magnetization in weak spin-orbit coupling limit.

Here we show that similar dynamics also occurs in the non-degenerate gas, which is much more readily accessible in experiments. Furthermore, we obtain analytic expressions for the total magnetization for \textit{arbitrary} values of the spin-orbit coupling. 

Rotating the spin polarized state to the diagonal basis, one finds that the density matrix has both diagonal and off-diagonal matrix elements ($A_{+} = A_{-} = 1/2$ and $A_{+-} = -\frac{1-i}{2\sqrt{2}}$), whose dynamics is given by Eqns.~(\ref{diagansatz}, \ref{offdiageq}). Transforming back to the hyperfine basis and integrating over momentum, the longitudinal magnetization density takes the form:
\begin{eqnarray}\label{magmagdensa}
m_{z}(\tilde r_{+},\tilde t) \sim  e^{-\frac{1}{2}\eta^2 \tilde r_{+}^2 - \frac{\tilde\alpha^{2}}{\eta^{2}}(1- \cos(2\tilde t))}\times \\\nonumber\cos(2 \tilde r_{+}\tilde\alpha (1 - \cos(\tilde t))\\\nonumber
\end{eqnarray}
and the total magnetization is:
\begin{equation}\label{totmag}
M(\tilde t) \sim \sqrt{\frac{2\pi}{\eta}}e^{-\frac{4\tilde\alpha^{2}}{\eta^{2}}(1- \cos(\tilde t))} \hspace{20mm}
\end{equation}
where we have ignored an overall normalization factor resulting from integration over momentum. The expression for the transverse magnetization density is rather cumbersome, but the total transverse magnetization remains zero at all times. 

From Eq.~\ref{totmag}, it is clear that the total longitudinal magnetization exhibits collapse and revival dynamics in time with a period which depends \textit{only} on the trapping potential, and is completely independent of the temperature or the spin-orbit coupling strength. At fixed temperature, for weak spin-orbit coupling $\tilde\alpha \ll 1$, our expression reads $M \sim 1 -4\tilde\alpha^{2}/\eta^{2}(1-\cos(\tilde t))$ \cite{stanescu}. In this limit, the magnetization exhibits sinusoidal oscillations with an amplitude which is proportional to $8\tilde\alpha^{2}/\eta^{2}$. For strong spin-orbit coupling, $\tilde\alpha \gg 1$, the magnetization becomes strongly peaked near $t = 2\pi n/\omega_{r}$ where $n$ is an integer, and decays exponentially, away from these points. 

The collapse and revival of the total magnetization is a trap effect. In a homogeneous system,  the momentum dependent spin-orbit magnetic field will simply cause the spins to dephase, particles with different momenta will precess at different rates, and the total magnetization will go to zero irreversibly on a timescale set by the spin-orbit coupling strength \cite{sherman}. It is also important to emphasize that the collapse and revival in the total magnetization described above has a different origin from what is observed in the experiments of Wang \textit{et al.} \cite{chinasoc}. We will discuss this in more detail later.

To understand the origin of the magnetization oscillations, we now turn to the dynamics of the longitudinal  and transverse magnetization density following the ramp. In a trapped geometry, from Eq.~\ref{magmagdensa}, we find that the longitudinal magnetization density exhibits periodic oscillations in space and in time. The temporal oscillations have a period of $t = 2\pi/\omega_{r}$, while at $t= \pi/\omega_{r}$ the spatial oscillations have a characteristic wave-length of $\lambda_{\text{sh}} = \pi~r_{\text{trap}}/(2\tilde\alpha)$ where $r_{\text{trap}} = \sqrt{\hbar/m\omega_{r}}$.

In Fig.~\ref{nonintfig} we plot the magnetization density normalized to the initial magnetization at the center ($m(r=0, t=0)$) as a function of time for the parameters above. Brighter colors indicate positive magnetization while darker colors indicate negative values of $m_{z}$. We choose a rather weak spin-orbit coupling strength in order to enhance the wavelength of the spatial oscillations at $t = \pi/\omega_{r}$. The transverse components of the spin are indicated by arrows whose length corresponds to the magnitude of the spin vector in the $x-y$ plane. At $t=0$, all spins are pointing in the $\uparrow$ direction indicated by the bright region in the density plot. Over time a transverse component develops and a spin helix emerges. At $t=\pi/\omega_{r}$, the spin oscillations reach the maximum amplitude proportional to $e^{-\frac{1}{2}\eta^{2} \tilde r_{+}^{2}}$, with a wave-length of $\Lambda_{\text{sh}}/r_{\text{trap}} =  \pi/2\tilde\alpha$. 

Energy conserving dynamics in phase space implies that the momentum of a particle is linked to its position. Moreover, the spin of an atom is linked to its momentum via the spin-orbit coupling. The spin-orbit Hamiltonian shifts the minimum of the dispersion to finite momenta. In a harmonically confined system, the iso-energy contours are circles in phase space, that are now shifted to finite momenta due to the spin-orbit coupling. A wave-packet polarized in the $\uparrow$ direction centered at $r = p = 0$ will follow the iso-energy contours in phase space, while simultaneously rotating in spin space. At $t=\pi/\omega_{r}$ the atomic wave-packet is centered around $r = 0$ in real space, and in order to conserve the total energy, the distribution will be centered around $\tilde p_{+} = \pm2\tilde\alpha\eta$ in momentum space. As a result, atoms with opposite momenta precess in opposite directions in spin space, producing a spin helix. At $t = 2\pi/\omega_{r}$, the atoms return to their original distribution in real and momentum space, and the initial state is recovered. The spin helix has a smaller net magnetization as compared to the fully polarized initial state, thus explaining the oscillations in the net magnetization. 

Experimentally, the spin density can be imaged \textit{in situ} using phase contrast imaging \cite{zwierleinps, sadler}. The wave-length of the spin helix is set by the spin-orbit interaction which is determined by the wave-length of the Raman beams. For the parameters used in the experiment of Cheuk \textit{et al.}, $\Lambda_{\text{sh}} \sim 0.5\mu$m \cite{zwierleinsoc}, which may be below the experimental resolution. However, the wave-length can be increased by decreasing the spin-orbit coupling strength. At present, experiments on spin-orbit coupled cold gases suffer from extremely short lifetimes ($< 500$ms) due to the large heating rates resulting from inelastic light scattering from the Raman beams \cite{spielmanthy}.  For typical trapping potentials, the timescale for the appearance of the spin helix is $t = \pi/\omega_{r} \sim 50$ms, so the short lifetimes may not be a major limitation in observing the spin helix.


\subsection{$\Omega_{R} \neq 0$}

We now turn to the experimentally relevant case where upon suddenly switching on the Raman beams, the atoms also experience a constant, momentum independent magnetic field, parametrized by a dimensionless parameter $\tilde B = \Omega_{R}/\omega_{r}$. In this limit, the problem cannot be solved analytically as the Zeeman and spin-orbit terms in Eq.~\ref{boltzmann} do \textit{not} commute.  Instead we numerically integrate the Boltzmann equation on a $4$D grid in phase space. We choose a $20 \times 20$ lattice in $\textbf{R}$ and $\textbf{p}$ with a spatial resolution of $\delta r = 2~r_{\text{trap}}$, where $r_{\text{trap}} = \sqrt{\hbar/m\omega_{r}}$ is the characteristic length scale of motion in the trap, and momentum resolution of $\delta p = 0.6~2\pi/\Lambda_{\text{th}}$. The integration is done using a split-step method that conserves total particle number and the total energy to high accuracy for sufficiently small time steps.

\begin{figure}
\begin{picture}(100, 200)
\put(-80, 20){\includegraphics[scale = 0.36]{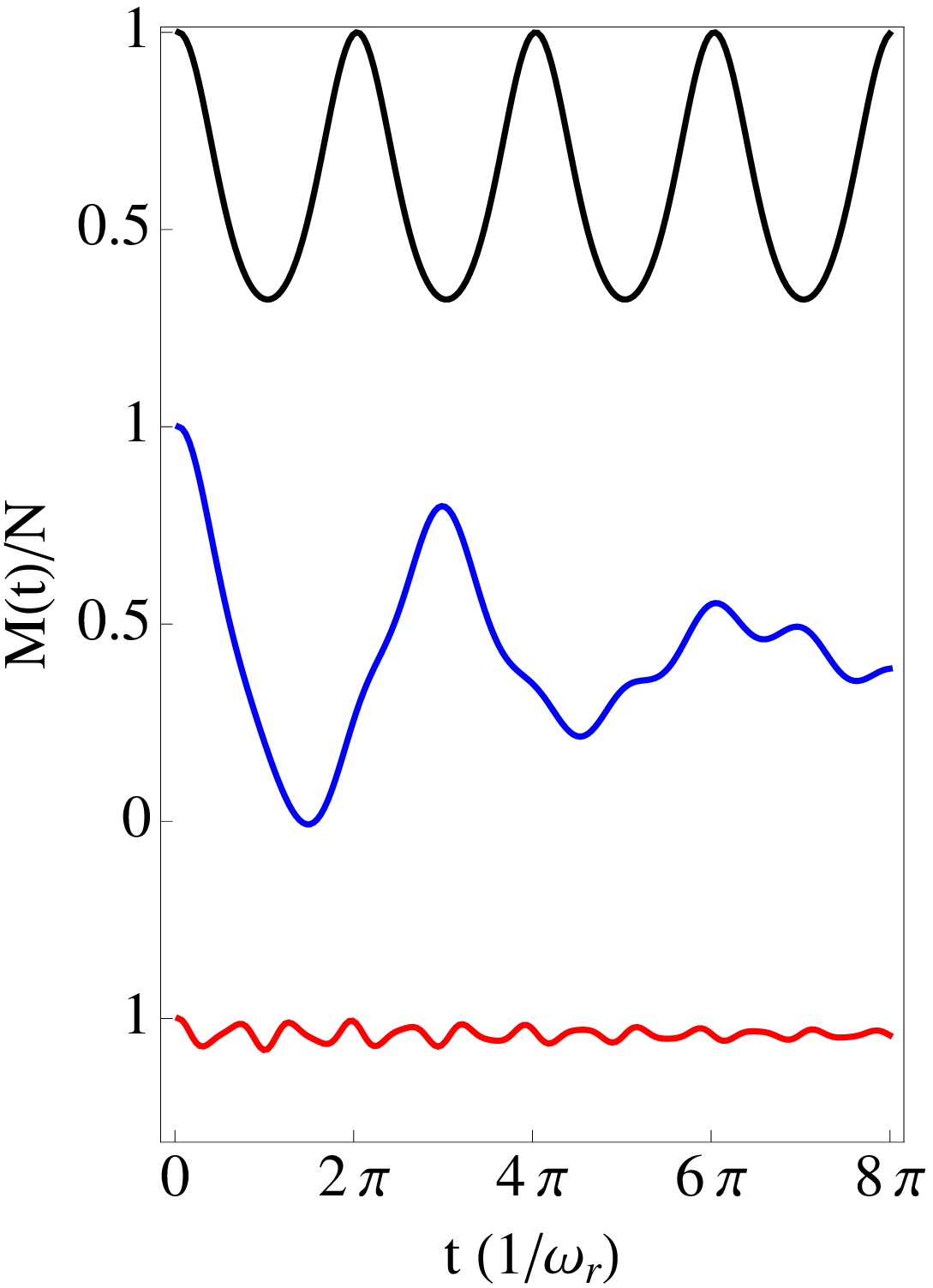}}
\put(40, 130){\includegraphics[scale=0.37]{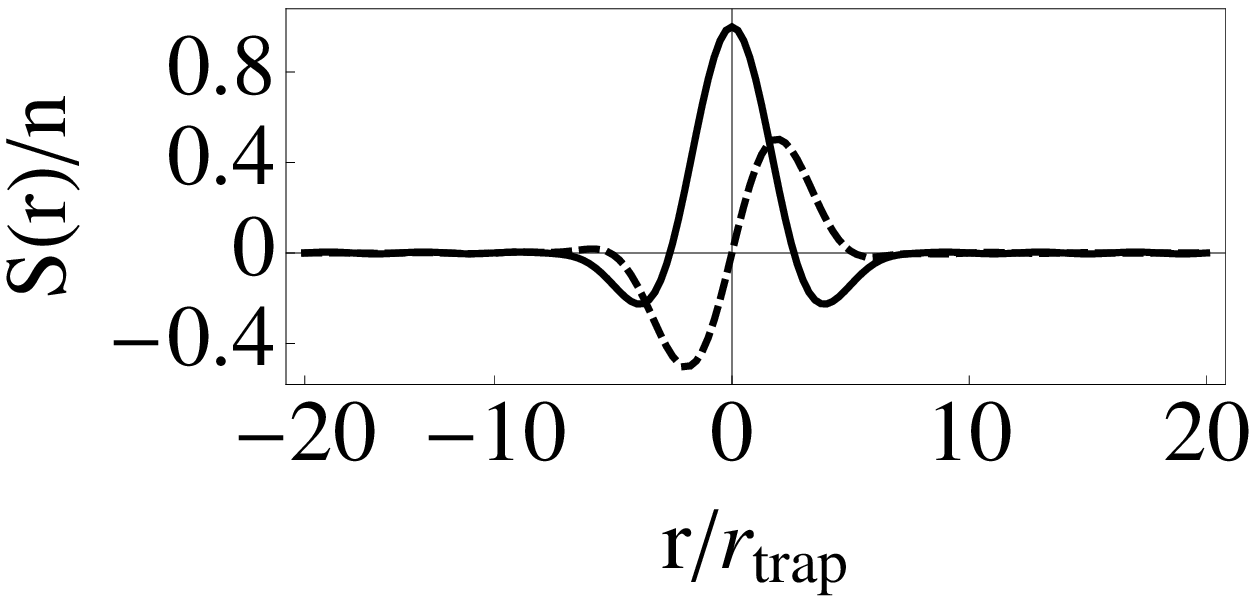}}
\put(40, 63){\includegraphics[scale=0.37]{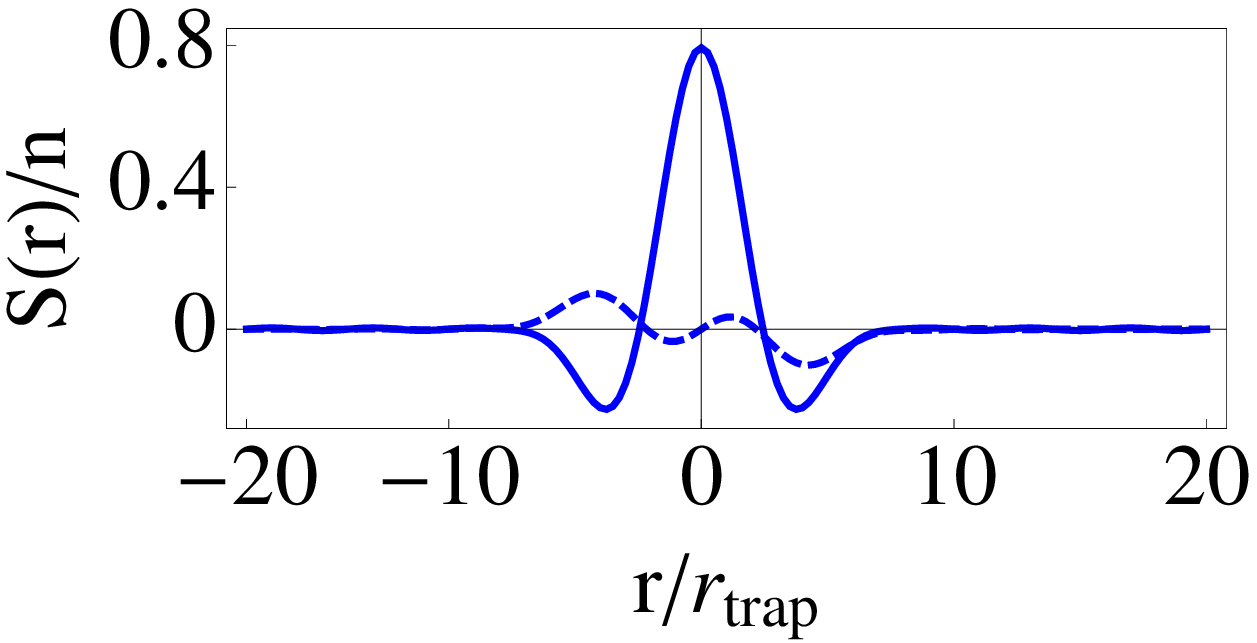}}
\put(40, -10){\includegraphics[scale=0.37]{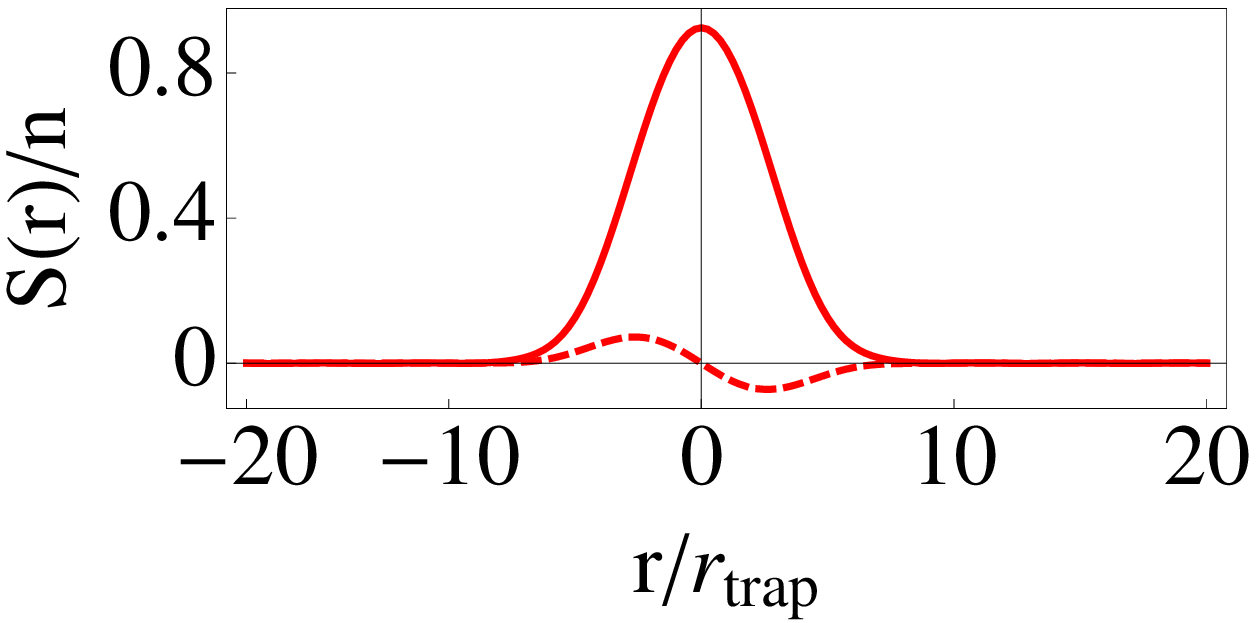}}
\end{picture}
\caption{\label{nonintdynfield}(Left) Time Evolution of the total magnetization in the system (see Eq.~\ref{magmagdens}) normalized to the total particle number for three different values of the dimensionless Zeeman coupling $\tilde B = \Omega_{R}/\omega_{r}$. In each case, we fix $\tilde\alpha = \sqrt{2}\alpha \sqrt{m/\hbar\omega_{r}} = 0.25$. From top to bottom: (Black) $\tilde B = 0$; (Blue) $\tilde B = 0.25$; (Red) $\tilde B = 1$. (Right) Magnetization density along the $r_{+} = x+y$ direction for the same values of the spin-orbit coupling strength and Zeeman field as in the left figure at fixed time $t = \pi/\omega_{r}$. The magnetization densities are normalized to the total central density denoted $n = n(r = 0, t=0) = \int d\textbf{p}[f_{\uparrow\uparrow}(\textbf{p}, 0, 0) + f_{\downarrow\downarrow}(\textbf{p}, 0, 0)]$. Solid curves represent the magnetization density in the $z$ direction (Eq.~\ref{magmagdens}), while the dashed curves indicate the magnetization density in the $x$ direction: $m_{x}(\textbf{r}, t = \pi/\omega_{r}) = \int d\textbf{p}~[f_{\uparrow\downarrow}(\textbf{p}, \textbf{r}, \pi/\omega_{r}) + f_{\downarrow\uparrow}(\textbf{p}, \textbf{r},\pi/\omega_{r})]$.}
\end{figure}

For simplicity we choose a fully polarized initial state, which is stationary in the absence of interactions or spin-orbit coupling. We then consider two limits, upon switching on the Raman beams: $\tilde B\sim \tilde\alpha$ and $\tilde B > \tilde\alpha$. The two parameters can be controlled independently as the magnitude of $\tilde\alpha$ is set by the wave-length of the Raman beams, and $\Omega_{R}$ is set by the laser intensity. The resulting dynamics of the total magnetization as well as the magnetization density is plotted in Fig.~\ref{nonintdynfield}.

As shown in Fig.~\ref{nonintdynfield}, the addition of a Zeeman term causes the magnetic field oscillations to decay over time. At long times, the total magnetization acquires a new steady state value which is smaller than $1$. As the strength of the Zeeman field is increased, the fully polarized initial state becomes increasingly stable, and the total magnetization remains close to $1$ at long times with small oscillations. 

The dynamics can be understood as follows: in the presence of a Zeeman field, each spin precesses about a new magnetic field, which is the sum of the spin-orbit magnetic field and the Zeeman field, and is tilted away from the x-y plane by an angle $\sin(\theta_{p}) = \tilde B/\sqrt{\tilde B^{2} + (\tilde\alpha \eta\tilde p_{+})^{2}}$. As the initial state has a Gaussian distribution of atoms with different momenta with spins pointing in the $z-$direction, atoms with momenta greater than $p > \tilde p_{\text{crit}} = \tilde B/\tilde \alpha\eta$ will primarily experience the spin-orbit magnetic field and precess about the $x-y$ plane, while atoms with momenta $p < p_{\text{crit}}$ will predominantly see the Zeeman field, and precess about the $z-$ axis. 

In a thermal gas, the characteristic width of the momentum distribution is set by $p_{\text{th}} = \sqrt{2\pi}\hbar/\Lambda_{\text{th}}$. If the Zeeman field is weak compared to the spin-orbit strength ($\alpha p_{\text{th}} >> \hbar\Omega_{R}$), the majority of the atoms still experience the spin-orbit magnetic field, and the spin density wave is preserved (as shown in the blue curves in Fig.~\ref{nonintdynfield}) for the first few oscillations. On longer timescales the magnetic field affects the spin precession of even the atoms with $p >> p_{\text{crit}}$ and  the spin density wave disappears. On very long times the system settles into a new steady state with a lower net magnetization. 

On the other hand, if the Zeeman field is strong ($\alpha p_{\text{th}}  << \hbar\Omega_{R}$), the majority of the atoms experience a net magnetic field which is only slightly tilted away from the $z$-axis.  As the initial state is fully polarized, the total magnetization remains close to $1$ at all times with rapid oscillations whose frequency is given by $\Omega_{R}$. The transverse spin density remains small at all times (see the dashed red curve in Fig.~\ref{nonintdynfield}) and no spin helix is found. 

The spin helix is the result of the interplay between dynamics in spin space and dynamics in phase space. Observing the spin helix therefore requires that the timescales for spin dynamics be comparable to the timescales for dynamics in phase space. 

In the presence of a large Zeeman field, as in the experiments of Wang \textit{et al.} \cite{chinasoc} ($E_{Z} \sim$kHz $\gg \hbar\omega_{r} \sim 50$Hz), the timescale for spin dynamics is set by the Zeeman term, which is much too short for any dynamics to occur in phase space. Thus while the total magnetization shows collapse and revivals (on a timescale set by $t \sim 2\pi/E_{Z} \sim 0.1$ms), the magnetization density in real space is unaffected. By contrast the collapse and revival dynamics discussed in Sec. III A occurs because the initial state is out of equilibrium in \textit{phase space}, when the Raman coupling is switched on. As a result, the magnetization oscillates on a timescale set by $t \sim 2\pi/\omega_{r} \sim 10$ms, and a spin spiral appears.

\section{Interactions} We now consider the effect of weak interactions on the dynamics discussed above. As discussed previously, we work in the the collisionless limit (sometimes referred to as the Knudsen regime), which is valid as long as $\tau_{\text{mf}} \ll \tau_{\text{coll}}$. 

Spin waves in dilute, quantum gases were studied theoretically by Bashkin and others \cite{bashkinspin, laloe, mullin, castaing} and observed in experiments on spin polarized hydrogen \cite{johnsonspin}. More recently, Du \textit{et al.} explored similar physics in a weakly interacting, thermal gas of $^{6}$Li atoms \cite{thomasspinwave}. Here we explore spin waves in a collisionless thermal gas with spin-orbit coupling. 
 
To compare with the results of Sec III,  we consider a fully polarized initial state which is the stationary state of the Boltzmann equation in the absence of spin-orbit coupling or interactions.  We then simultaneously  switch on the interactions, and the spin-orbit coupling. We define a dimensionless interaction strength $\tilde g = g^{\text{2D}}n/\hbar\omega_{r}$, where $n$ is the total initial density at the trap center $n = n(r = 0, t=0) = \int d\textbf{p}[f_{\uparrow\uparrow}(\textbf{p}, 0, 0) + f_{\downarrow\downarrow}(\textbf{p}, 0, 0)]$. For typical trapping potentials used in experiments, $\omega_{r} \sim 2\pi\times100$ Hz, the Knudsen regime corresponds to $\tilde g \sim 0.1$ \cite{thomasspinwave}. We choose $\tilde g = 0.25$. Furthermore, as shown in Sec III B, a large Zeeman field ($\tilde B \gg \tilde\alpha$) stabilizes the spin polarized state and produces little spin dynamics. Hence we consider $\tilde B = \tilde \alpha = 0.25$.

\begin{figure}
\begin{picture}(200, 200)
\put(-23, 80){\includegraphics[scale=0.28]{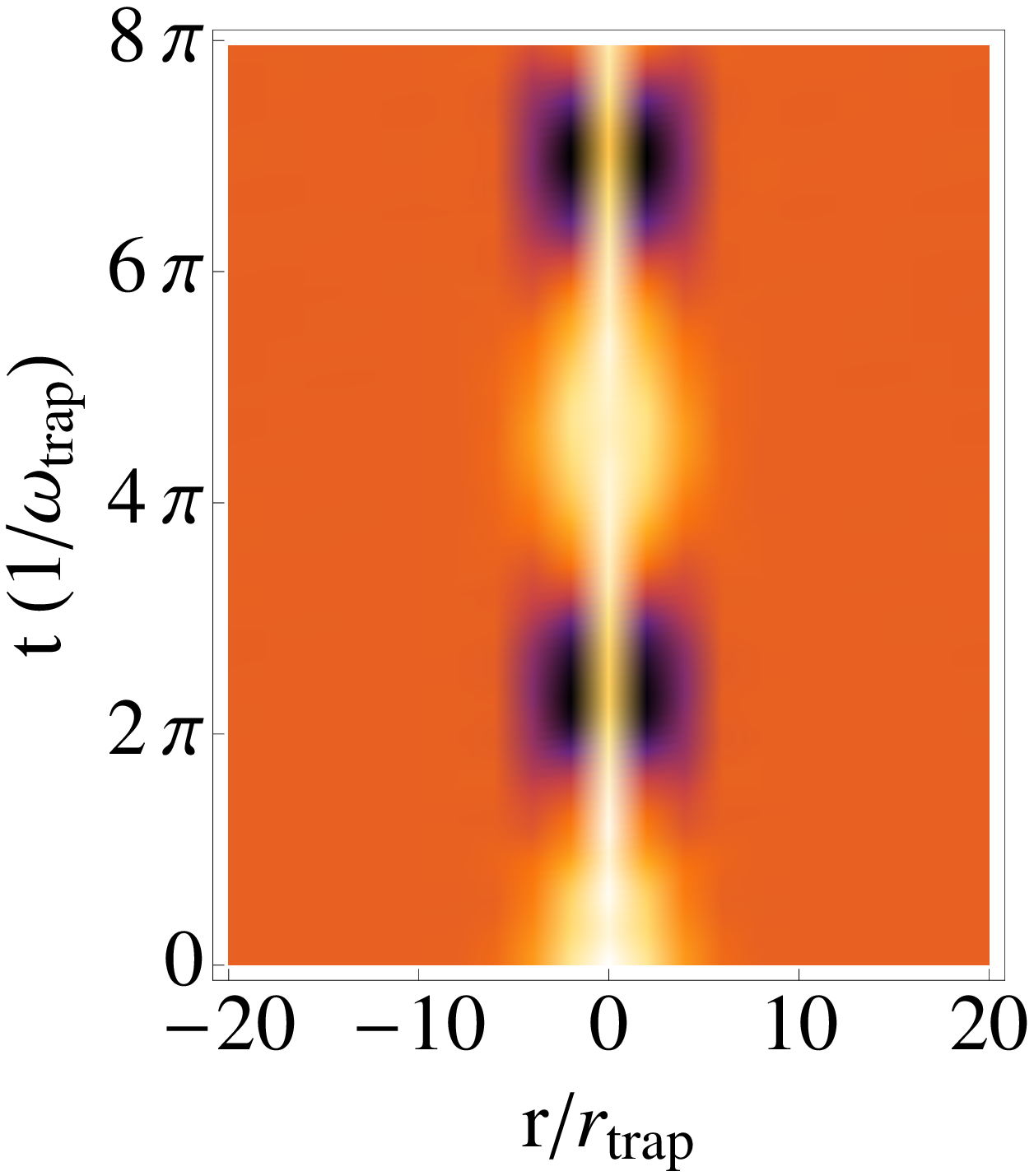}}
\put(90, 130){\includegraphics[scale=0.37]{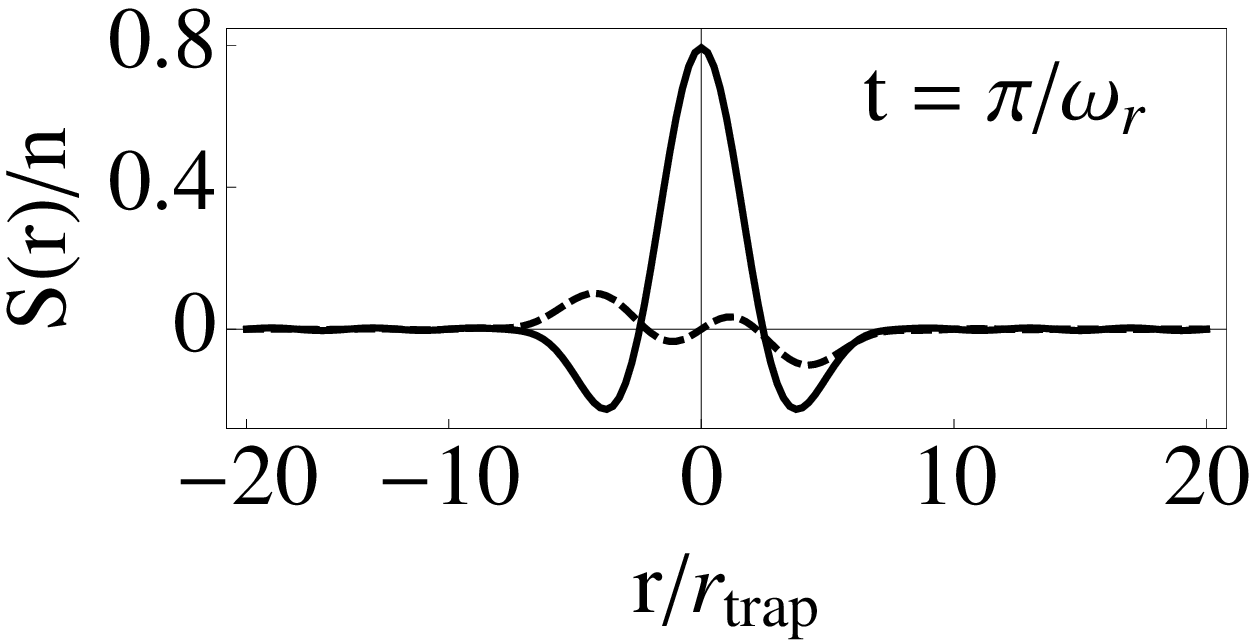}}
\put(90, 60){\includegraphics[scale=0.37]{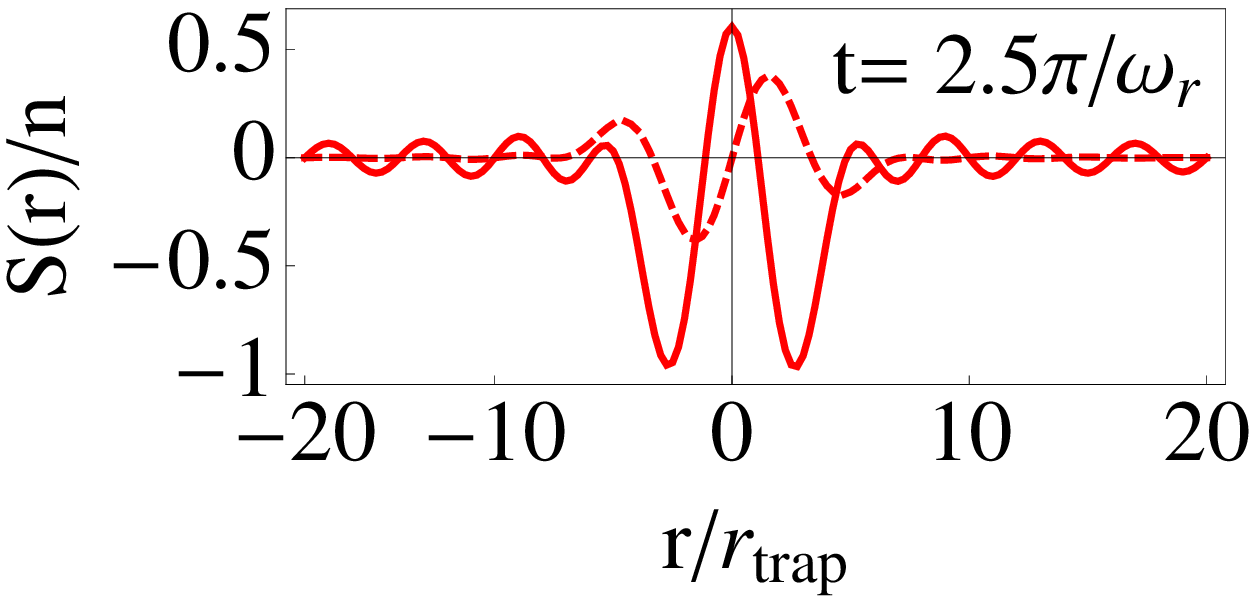}}
\put(90, -10){\includegraphics[scale=0.37]{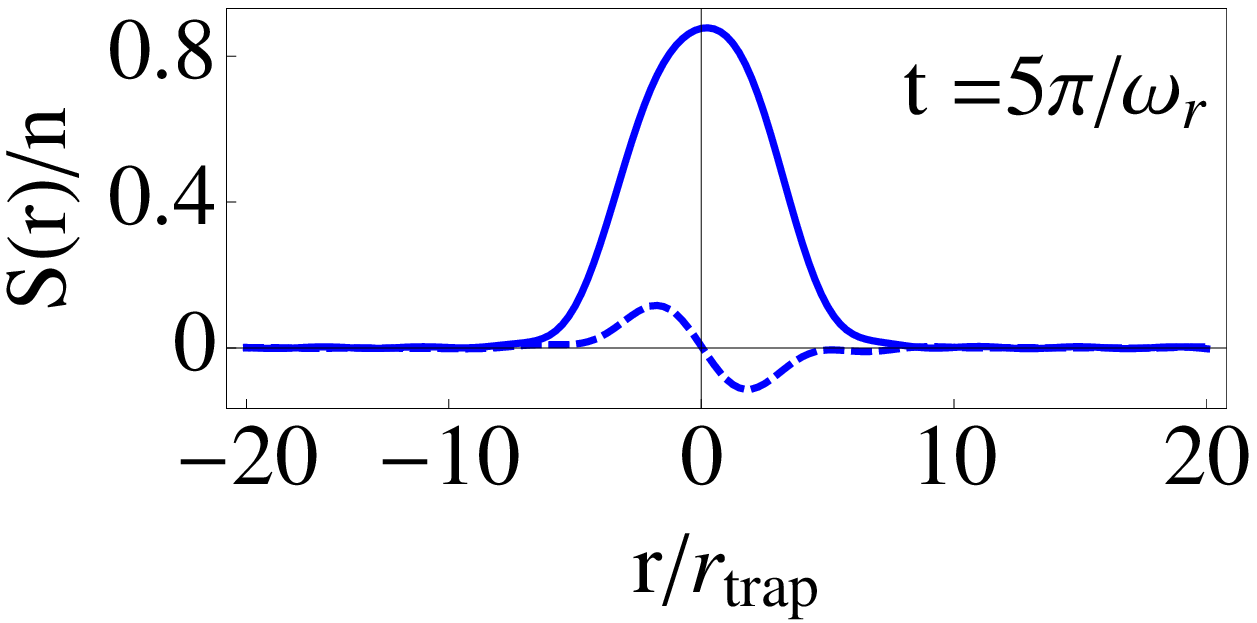}}
\put(-23, -10){\includegraphics[scale=0.28]{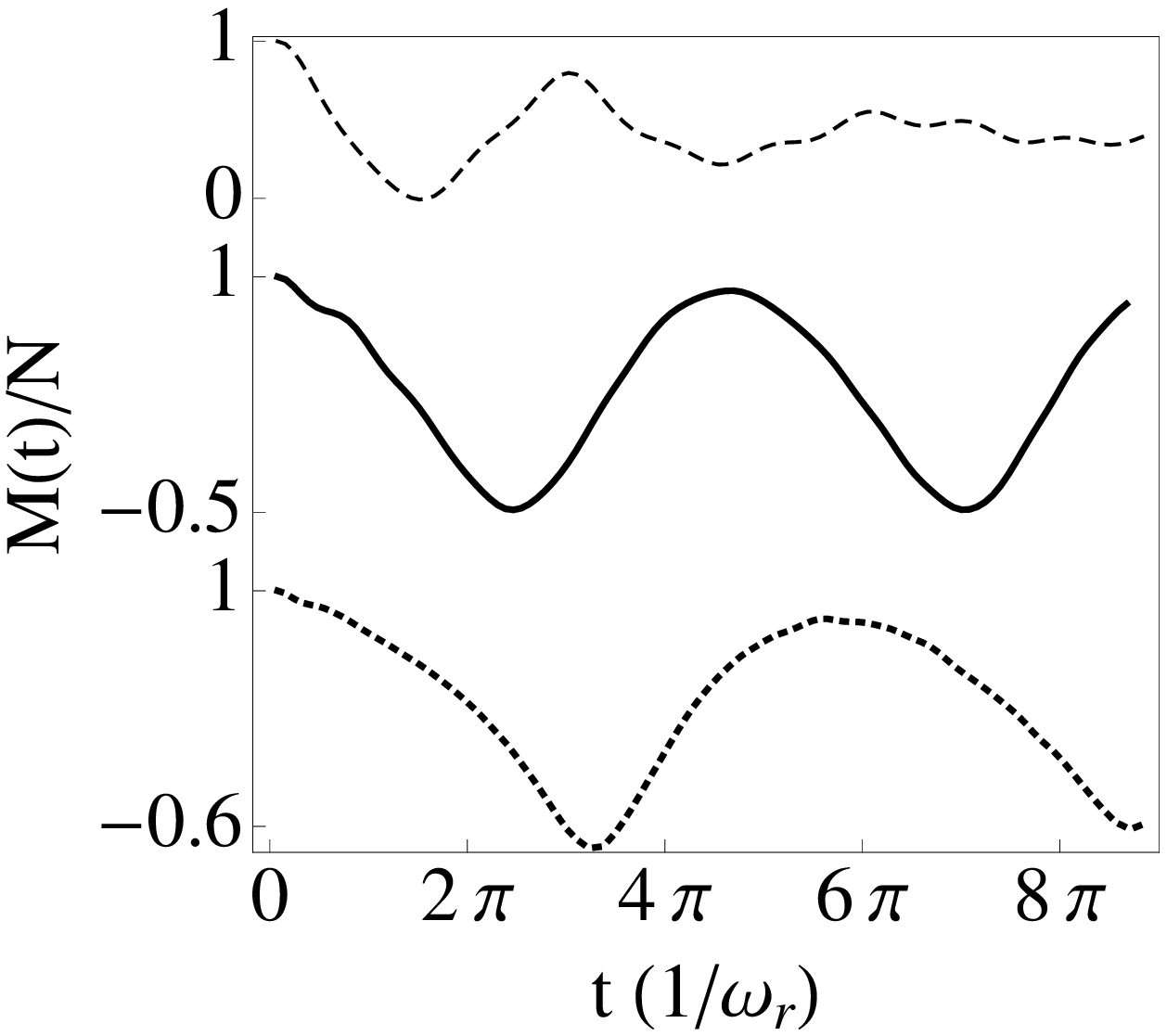}}
\end{picture}
\caption{\label{intfiga} Top Left: Density plot showing the magnetization density along $r_{+}$ (see Eq.~\ref{magmagdens}), normalized to the total initial central density as a function of time. The parameters are $\tilde g = \tilde B = \tilde \alpha = 0.25$ (Recall that $\tilde g =  gn/\hbar\omega_{r}$, $\tilde B = \Omega_{R}/\omega_{r}$ and $\tilde \alpha =  \sqrt{2}\alpha \sqrt{m/\hbar\omega_{r}}$. Brighter colors indicate positive magnetization while darker colors indicate negative magnetization. Bottom Left: Total magnetization in the entire system as a function of time for three different values of $\tilde g$. In each plot, $\tilde B = \tilde \alpha = 0.25$. From top to bottom $\tilde g = 0$ (dashed), $\tilde g = 0.25$ (solid) and $\tilde g = 0.5$ (dotted).  Panels on Right: Longitudinal (solid) and transverse (dashed) magnetization densities in the $r_{+}$ direction at different times. The top panel shows the spin helix without interactions for $\tilde B = \tilde \alpha = 0.25$ (same as the central panel in Fig.~\ref{nonintdynfield}) for comparison. The central and bottom panels are snapshots of the longitudinal and transverse magnetization at time $t = 2.5\pi/\omega_{r}$ and $t = 5\pi/\omega_{r}$ for the same parameters as in the Top Left.}
\end{figure}

In Fig.~\ref{intfiga} we plot the total magnetization density along the $r_{+}$ direction and the total net magnetization as a function of time. As is apparent from the figure, the inclusion of interaction drives large magnetization oscillations on a characteristic timescale much larger than $t = \pi/\omega_{r}$. This timescale grows as the interaction strength is increased. In real space, the magnetization minimum is manifested as a large amplitude longitudinal spin wave. The panels on the right show the magnetization density in the longitudinal and transverse directions (solid and dashed respectively) at two different times $t = 2.5\pi/\omega_{r}$ (red) and $t = 5\pi/\omega_{r}$ (blue) for parameters corresponding to the density plot. The spin densities for the spin helix (c.f Fig.~\ref{nonintdynfield}(right) center panel) for the same values of $\tilde B$ and $\tilde \alpha$, but $\tilde g = 0$ are shown in the top panel for comparison.  The amplitude of the spin wave seen here is much larger than what was observed in previous studied of the spin-$\frac{1}{2}$ gas in the absence of spin-orbit coupling \cite{thomasspinwave}. 

Interactions alter the physics of the non-interacting gas in two crucial ways.  Forward scattering modifies the iso-energy contours in phase space:  spin $\uparrow$ atoms experience a mean-field proportional to the density of the $\downarrow$ atoms and vice versa. More importantly, due to the exchange interaction, when two atoms collide, they precess about the common axis of their total spin. As argued by Lhuillier and Lalo\"e, it is this effect that gives rise to spin waves in the collisionless gas \cite{laloe}. In the experiments of Du \textit{et al.} \cite{thomasspinwave}, the initial state was polarized along the $x-$direction, and spin dynamics was the result of the negligible difference in the trapping potentials experienced by the spin $\uparrow$ and $\downarrow$ atoms.  Consequently the amplitude of the resulting spin wave was very small, $m_{z}(\textbf{r})/n_{0} \ll1$ \cite{thomasspinwave}. 

By contrast in the present case the initial state is polarized along $z$. Absent spin-orbit coupling, this state has \textit{no} dynamics as it is stationary with respect to the Zeeman field and the interaction term. spin-orbit coupling however causes the spins to precess about the $x-y$ plane with a precession rate proportional to the momentum. Over time, atoms with different momenta precess at different rates and can collide with one another. The atomic collisions subsequently lead to a dynamical spin segregation in real space \cite{laloe}. Unlike the experiments of Du \textit{et al.}, the initial spin precession in our system occurs due to spin-orbit coupling, which is not a small effect. Consequently the amplitude of the resulting spin wave is also larger than what was found in Ref.~\cite{thomasspinwave}.

We remark that the amplitude of the total magnetization oscillations, and the associated spin texture in real space, depends non-monotonically on the strength of the Zeeman field. For $\tilde B \ll \tilde\alpha$, and weak interactions $\tilde g \ll 1$, the magnetization oscillations are only slightly larger than those studied in Sec III A, and the real space magnetization density resembles the spin helix shown in Fig.~\ref{intfiga} (top right). The magnetization oscillations reach a maximum at some $\tilde B_{\text{crit}}$, which depends on $\tilde g$ and $\tilde \alpha$. Upon further increase of $\tilde B$, the spin $\uparrow$ state becomes stable and there is no spin dynamics.

\section{Summary and Conclusions} 

In summary, we studied the spin dynamics in a weakly interacting, non-degenerate Fermi gas following a sudden ramp of the spin-orbit coupling strength. In the absence of interactions and a Zeeman field, we produced analytic expressions for the total magnetization and the magnetization density for arbitrary values of the spin-orbit coupling, trap frequency and the temperature. We argued that a fully polarized initial state will give rise to a spin helix on a timescale set by the trapping period with a wave-length that depends on the ratio of the spin-orbit interaction to the trap frequency. For the high temperature gas, we generalized the analytic results obtained by Stanescu \textit{et al.} \cite{stanescu} to arbitrary spin-orbit coupling strength.

We then numerically studied the spin dynamics in the presence of interactions and a Zeeman field, highlighting the role played both separately and collectively by these terms. For weak Zeeman fields the spin helix is preserved but it disappears for very large Zeeman fields, as the fully polarized initial state becomes increasingly stable. In the presence of interactions however the dynamics is more complicated. Interactions tend to \textit{enhance} the amplitude of the spin oscillations. We explain this effect as a dynamical spin segregation driven by exchange.   

Finally, we briefly comment on the role of Fermi statistics in our calculations. In principle, all the results discussed in this paper apply equally well to bosons. The spin helix discussed in Sec III is a single-particle effect, and should also occur in a bosonic system. Importantly, it is a \textit{dynamical} effect and should be contrasted with the spin stripe phase predicted by Ho and Zhang, which occurs in equilibrium spin-orbit coupled Bose-Einstein condensates \cite{jasonspinorbit}. 

To conclude, our work shows that a seemingly one-particle term in the Hamiltonian (i.e. spin-orbit coupling), when coupled with coherent non-equilibrium dynamics of the Fermi gas, could manifest rather non-obvious and complex (and more importantly, observable) behavior in cold atoms and molecules even at relatively high temperature where the quantum degeneracy of the system is not of any key importance. The dynamical physics we predict here is unlikely to manifest itself in any solid state systems (and can only be seen in atomic gases) although the basic concept of spin-orbit coupling originates in solids as discussed in detail in Refs.~\cite{hirsch, zutic}.

\section{Acknowledgements} 

It is a pleasure to thank Juraj Radic for his insights, and his careful reading of the manuscript.  We also gratefully acknowledge discussions with Victor Galitski, I.V.Tokatly and E. Y. Sherman. This work is supported by JQI-NSF-PFC, AFOSR-MURI, and ARO-MURI.

\end{document}